\documentclass[prl,aps,twocolumn,showpacs,floatfix]{revtex4-1}
\usepackage{mathrsfs}
\usepackage{color}
\usepackage{graphicx}
\usepackage{dcolumn}
\usepackage{bm}
\usepackage{hyperref}
\usepackage{hypernat}
\usepackage[latin1]{inputenc}
\usepackage{pstricks}
\usepackage{amsmath,amssymb}
\usepackage{version}
\usepackage{epstopdf}
\usepackage{ulem}

\DeclareGraphicsExtensions{.eps,.ps}

\begin{document}

\preprint{APS/123-QED}
\title{Multistage Coupling of Laser-Wakefield Accelerators with Curved Plasma Channels}

\author{J. Luo$^{1,7}$}
\author{M. Chen$^{1,7}$}
\email{minchen@sjtu.edu.cn}
\author{W. Y. Wu$^{1,7}$}
\author{S. M. Weng$^{1,7}$}
\author{Z. M. Sheng$^{1,2,5,6,7}$}
\email{z.sheng@strath.ac.uk}
\author{C. B. Schroeder$^{3}$}
\author{D. A. Jaroszynski$^{2,6}$}
\author{E. Esarey$^{3}$}
\author{W. P. Leemans$^{3}$}
\author{W. B. Mori$^{4}$}
\author{J. Zhang$^{1,7}$}
\affiliation{
$^1$Key Laboratory for Laser Plasmas (MOE), School of
Physics and Astronomy, Shanghai Jiao Tong University, Shanghai 200240, China\\
$^2$SUPA, Department of Physics, University of Strathclyde, Glasgow G4 0NG, UK\\
$^3$Lawrence Berkeley National Laboratory, Berkeley, CA 94720, USA\\
$^4$University of California, Los Angeles, CA 90095, USA\\
$^5$Tsung-Dao Lee Institute, Shanghai Jiao Tong University, Shanghai 200240, China\\
$^6$Cockcroft Institute, Sci-Tech Daresbury, Cheshire WA4 4AD, UK\\
$^7$Collaborative Innovation Centre of IFSA (CICIFSA),
Shanghai Jiao Tong University, Shanghai 200240, China
}
\date{\today}

\begin{abstract}
Multistage coupling of laser-wakefield accelerators is essential to overcome laser energy depletion for high-energy applications such as TeV level electron-positron colliders. Current staging schemes feed subsequent laser pulses into stages using plasma mirrors, while controlling electron beam focusing with plasma lenses. Here a more compact and efficient scheme is proposed to realize simultaneous coupling of the electron beam and the laser pulse into a second stage. A curved channel with transition segment is used to guide a fresh laser pulse into a subsequent straight channel, while allowing the electrons to propagate in a straight channel. This scheme benefits from a shorter coupling distance and continuous guiding of the electrons in plasma, while suppressing transverse beam dispersion. With moderate laser parameters, particle-in-cell simulations demonstrate that the electron beam from a previous stage can be efficiently injected into a subsequent stage for further acceleration, while maintaining high capture efficiency, stability, and beam quality.
\end{abstract}

\maketitle
Laser wakefield accelerators (LWFAs) have attracted considerable attention in recent years as a promising new accelerator technology \cite{Tajima1979PRL,  Pukhov2002APB, Esarey2009RMP, Martins2010NP}. They are capable of supporting enormous acceleration gradients, as high as hundreds of GeV/m. This makes it possible to build compact accelerators in university-scale laboratories for many applications, such as compact electron diffraction devices \cite{He2013APL}, high-energy particle accelerators \cite{Karsch2007NJP, Kim2013PRL, Wang2013NC}, and tabletop radiation sources \cite{Corde2013RMP, Yan2017NP, Cipiccia2011NP, Luo2016SR}.

Among the many applications, perhaps the most intriguing, and challenging application is a LWFA-based TeV level electron-positron linear collider \cite{Leemans2009PT, Schroeder2010PRSTAB}. However, the energy gained by electrons in a single-stage LWFA is limited by several effects, including electron dephasing, laser diffraction, and laser energy depletion. Although a few schemes have been proposed to increase the single-stage energy gain by mitigating dephasing and diffraction, a single-stage LWFA is still limited by pump depletion. Currently, single stage acceleration up to 10 GeV energy is thought to be a reasonable value, given present laser technology and plasma density scaling, and worldwide effort is pursuing this goal \cite{Leemans2014PRL, Lu2007PRSTAB, Kneip2009PRL, Clayton2010PRL, Liu2011PRL, Guillaume2015PRL, Mirzaie2015SR}. For future TeV colliders, multistage coupling of LWFAs is inevitable. Electron beams accelerated in a first stage should be injected into a second wakefield stage that is driven by a fresh laser pulse. Due to the micrometer size electron beams and similar size wake structures, the synchronization precision at femtosecond scale, and the limited coupling distance, multistage coupling is considered to be very challenging. Recently, staged acceleration was initially demonstrated by Steinke \textit{et al.} \cite{Steinke2016N}. By using plasma mirrors \cite{Sokollik2010AIPCP} to reflect a fresh laser pulse and using a plasma lens \cite{Tilborg2015PRL} to refocus electrons, about 3.5\% of the electron beam charge was coupled into the second stage, which produced 100 MeV energy gain. In this staging layout, both the dedicated plasma mirror and lens had to be installed between the LWFA stages. Besides its complexity, the matching of the electron beam between stages is still very challenging in achieving efficient coupling. The coupling efficiency must be near 100\%, if one considers the requirements for up to a hundred stages. Thus, a simple and efficient multistage coupling scheme is highly desirable.

Besides plasma mirrors, bending plasma channels may be considered to guide lasers \cite{Ehrlich1996PRL, Reitsma2007POP, Chen2016LSA, Palastro2017POP}. Reitsma \cite{Reitsma2007POP} theoretically studied laser propagation in curved plasma channels and found an equilibrium laser centroid trajectory. Chen \cite{Chen2016LSA} and Palastro \cite{Palastro2017POP} discussed applications of synchrotron radiation based on these curved plasma channels and one patent was published on compact undulators and radiation source based on curved channels \cite{Hooker2007P}. In this paper, we propose the use of a curved plasma channel to enable a compact multistage LWFA. Instead of plasma lenses and mirrors, a specially designed straight and bent plasma channel is used to simultaneously couple an electron beam and a laser pulse into a second stage. In this case, the electron beam propagates in a straight plasma channel, it is constrained to the channel by the wakefield. A transition curved plasma channel simultaneously guides a fresh laser pulse into the straight section, generating a new wakefield with the appropriate phase to continue accelerating the electrons. Particle-in-cell (PIC) simulations confirm that stable and efficient laser guiding and electron coupling can be achieved using this scheme.

\begin{figure}
\includegraphics[width=0.48\textwidth]{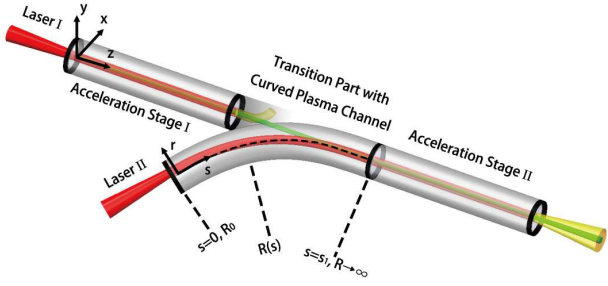}
\caption{
Schematic diagram of coupling the two LWFA acceleration stages via a curved plasma channel with trajectories of fresh lasers (red), depleted lasers (yellow) and electrons (green).
}
\label{fig1}
\end{figure}

A schematic view of this scheme is shown in Fig. \ref{fig1}. After driving plasma waves and depleting its energy in stage I, laser \uppercase\expandafter{\romannumeral1} is exhausted and deflected off the plasma in the connecting region. While a second fresh laser pulse (laser \uppercase\expandafter{\romannumeral2}) is transported from the entrance of the curved channel to the straight channel, electrons move along the straight channel and penetrate the wall of the curved channel, into the second straight section where they can be trapped for an appropriately timed laser pulse.

\begin{figure}
\includegraphics[width=0.48\textwidth]{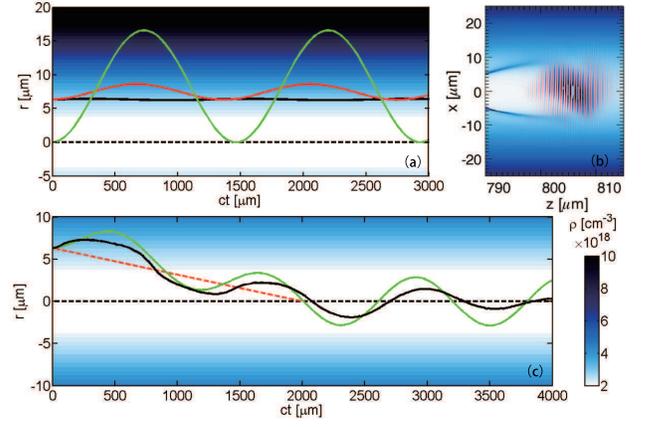}
\caption{
Laser centroid trajectories from TDSE and PIC simulation in a curved plasma channel with fixed radius of curvature (a) and the proposed transition curvature (c) with $\lambda_l=0.8$ $\rm \mu m$, $w_0=8$ $\rm \mu m$, $R_0=10$ $\rm mm$, $n_0=10^{-3}\ n_{cr}$. Dashed lines represent the centre of channels (black) and the laser equilibrium trajectory (red), respectively. A snapshot from a 2D-PIC simulation in (b) shows the laser profile for injection into a straight channel. All results account for relativistic laser intensity effects, except the black solid line in (a). 
}
\label{fig2}
\end{figure}

The key aspect of this scheme is stably and efficiently guiding of laser \uppercase\expandafter{\romannumeral2} into the straight channel without large energy loss. Moreover, laser centroid oscillation in the second acceleration stage should be avoided, which usually occurs if the laser is injected off-axis or obliquely into the straight channel \cite{Chen2016LSA}. We first focus on the optimal bending shape of the curved channel. For simplicity, we set the z axis to be along the straight channel centre. Provided the laser is linearly polarized in the y direction, the evolution of the laser pulse can be described by $(c^2{\nabla}^2-\partial^2/\partial t^2)A_y=\omega_{p}^{2}A_y$. The normalized $A_y$ is $eA_y/m_ec^2=a\cdot exp(ikz-i\omega_lt)/2+c.c.$,and $\omega_p=\sqrt{4\pi n_p e^2/m_e}$ is the plasma frequency. A transverse Gaussian pulse can stably propagate in a straight plasma channel with a parabolic transverse shape: $n_p (r)=n_0 + \Delta n \cdot (r^2/w_0^2)$, with $n_0=n_p(r=0), \Delta n=1.13\times 10^{20}(cm^{-3})/w_0^2(\mu m^2)$, $w_0$ the laser focal spot radius, and $r$ the radial distance to the centre axis of the channel \cite{Esarey2009RMP}. For a curved plasma channel with fixed radius of curvature $R$, it is convenient to introduce a co-moving coordinate $\xi=s-ct$, where s is the laser propagation distance along the channel centre. After applying the slowly-varying envelope and paraxial approximations, and keeping terms to lowest order in $r/R$, the laser envelope evolution equation can be expressed as
\begin{equation}
i\frac{\partial a}{\partial t} = [- \frac{c^2}{2\omega_l} \frac{\partial^2}{\partial r^2}+ \frac{\omega_l}{2} \frac{n_0}{n_{cr}} (1+\frac{\Delta n}{n_0} \frac{r^2}{w_0^2})- \omega_l \frac{r}{R}]a
\label{eq1}
\end{equation}
where $n_{cr}$ is the critical density for the laser \cite{Reitsma2007POP}. Equation \ref{eq1} takes the form of the Schr\"odinger equation, where the initial value is $a(t=0,r)=a_0exp[-(r-r_0)^2/w_0^2]$. The terms in the bracket on the right-hand side of Eq. \ref{eq1} correspond to a Hamiltonian operator. For a straight channel $R\rightarrow \infty$, the Hamiltonian is symmetric about $r=0$, therefore the incident laser can propagate stably if the initial laser centroid is on-axis (i.e. $r_0=0$). For a channel with a finite radius of curvature $R$, in a similar way, a laser beam can propagate without transverse oscillation when they are injected at transverse position $r_0 = r_{equ} = (n_{cr}/\Delta n)w_0^2/R$. To confirm this, Eq. \ref{eq1} has been numerically solved using a spectral fitting method, and the laser centroid trajectories ($r_c=\int{r|a|^2dr} / \int{|a|^2dr}$) are shown in Fig. \ref{fig2}(a). In this calculation, the channel curvature effects represented by the last term in Eq. \ref{eq1} are regarded as a potential. Therefore, the final laser centroid motion along the laser propagation is presented as in a straight channel. The black solid line in Fig. \ref{fig2}(a) shows the stable propagation trajectory of a laser with a transverse deviation distance from the channel centre of $r_{equ}=6.33$  $\rm \mu m$.

However, in Eq. \ref{eq1} the effects of relativistic laser intensities have been neglected. In our staging acceleration scheme, the incident laser intensity is $a_0=2$. Relativistic effects change the plasma refractive index and therefore cannot be neglected. In an improved calculation, the background plasma density $n_0$ in Eq. \ref{eq1} is replaced by $n_0/(1+|a|^2/2)^{1/2}$, which makes the Hamiltonian depend on the laser intensity and the equation take on the form of a time-dependent Schr\"odinger equation (TDSE). The recalculated laser centroid motion is shown by the red line in Fig. \ref{fig2}(a). As one can see, a slight centroid oscillation appears even when the laser is injected from a previous equilibrium position.

To avoid laser transverse oscillations in the second acceleration stage, which usually causes electron transverse loss and beam energy loss through betatron radiation, the laser centroid should be guided to the centre of the straight channel, which requires the injection angle to be as small as possible \cite{Esarey2009RMP}. With this constraint, the motion along the equilibrium position in the curved channel is not optimal. Actually, neither $r_0=0$ or $r_0=2\ r_{equ}$ are optimal, although the laser centroid can swing back to $r=0$ after a propagating a distance of integer times of $\Lambda_{os}/2 = \pi ^2 w_0^2/ \lambda_l$ \cite{Esarey2009RMP, Chen2016LSA}, represented by the green line of Fig. \ref{fig2}(a). In this case the tolerable exit region is very short due to the large transverse oscillation. More importantly, over this region the large plasma density difference along transverse direction leads to distortion of the laser profile, which increases its duration. These non-paraxial effects cannot be described by Eq. \ref{eq1}. A typical laser profile at the point where the laser enters the straight channel from a two-dimensional (2D) PIC simulation with $r_0=2\ r_{equ}$ is plotted in Fig. \ref{fig2}(b). Such severe distortion of the laser pulse reduces the stability and energy conversion efficiency in the second acceleration stage.

\begin{figure}
\includegraphics[width=0.48\textwidth]{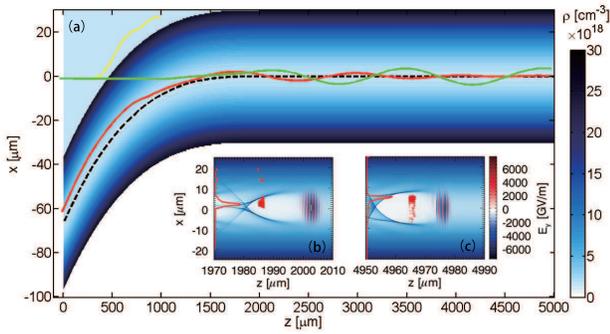}
\caption{
2D-PIC simulations of the multistage coupling scheme based on curved plasma channels. (a) Centroid trajectory of laser  \uppercase\expandafter{\romannumeral1} (yellow), laser \uppercase\expandafter{\romannumeral2} (red), and the electron beam (green). Insets (b) and (c) are snapshots of the electric field of laser \uppercase\expandafter{\romannumeral2}, plasma electron density, and electron beam (red points and red lines) at two propagation distances.
}
\label{fig3}
\end{figure}

To solve this problem, we propose using a variable curvature plasma channel. Considering the experimental feasibility, a variable curve characterized by fixing $(s_1-s)\cdot R^{\alpha}$ is chosen, where $s_1$ is the total length of the bent channel, as shown in Fig. \ref{fig1}. Thus, we find
\begin{equation}
r_{equ}=\frac{n_{cr}}{\Delta n} \frac{w_0^2}{R}=\frac{n_{cr}}{\Delta n} \frac{w_0^2}{R_0} (\frac{s_1-s}{s_1})^{1/\alpha}
\label{eq2}
\end{equation}
where $R_0$ is the curvature at $s=0$. The gradual decrease in $r_{equ}$ guides the laser centroid from the original equilibrium position to the channel centre, accompanied by oscillations in the direction perpendicular to $dr_{equ}/ds$. Thus, a fixed $dr_{equ}/ds$ is consistent with confining the oscillation amplitude, which results $\alpha=1$, and $(s_1-s) \cdot R=s_1 R_0$   is an appropriate selection for the transition curve. The tilt angle then is $\theta=(s_1-s)^2/2s_1 R_0$. By Taylor expanding $\theta$, keeping terms to lowest order in $s$, and taking $s\approx ct$ (i.e. regarding the pulse duration as negligible compared with the total length of the curve), the centre axis of the transition curve channel can be described as:
\begin{equation}
\left\{ \begin{array}{ll}
z =  \int{d(s_1-s) \cdot cos\theta} \approx s \approx ct\\
x =  \int{d(s_1-s) \cdot sin\theta} \approx (s_1-ct)^3/(6s_1R_0)
\end{array} \right.
\label{eq3}
\end{equation}
For such a curved channel the calculated equilibrium trajectory of the laser centroid (Eq. \ref{eq2}) is represented by the red dashed line in Fig. \ref{fig2}(c), with $s_1=2$ $\rm mm$. The theoretical prediction of Eq. \ref{eq1} is represented by the green solid line, which is consistent with the PIC simulation shown by the black solid line, apart from slight damping of the oscillation amplitude. The guiding centre trajectories are similar, which is consistent with our expectations. Furthermore, in this scheme, the laser remains in a lower plasma density region than that for the green line case in Fig. \ref{fig2}(a), which better preserves the laser quality in the transition region. A typical PIC simulation result of the laser profile when coupling into the straight channel is shown in Fig. \ref{fig3}(b), with initial laser parameters similar as the one shown in Fig. \ref{fig2}(b). As one can see, by using the transition curved channel, the laser profile is well maintained, which benefits the subsequent second stage acceleration.

We use PIC simulations to study multistage coupling for both the electron and laser beams. A typical 2D simulation result using OSIRIS code \cite{Fonseca2002LNCSE} is shown in Fig. \ref{fig3}, where we have chosen a laser of sin-squared longitudinally envelope with $a_{10}=0.7$, $w_{10}=8$ $\rm \mu m$, pulse duration $\tau_{10}=15$ $\rm fs$, as the exhausted laser \uppercase\expandafter{\romannumeral1}, a laser of Gaussian longitudinally envelope with $a_{20}=2.0$, $w_{20}=8$ $\rm \mu m$, $\tau_{20}=20$ $\rm fs$ as the fresh laser \uppercase\expandafter{\romannumeral2}, and a uniformly distributed pre-accelerated electron beam with $r_{b}=0.5$ $\rm \mu m$, $l_{b}=2.0$ $\rm \mu m$, initial energy $\rm E=1\ GeV$, $(\rm \Delta E)_{FWHM}=50\ MeV$, initial momentum $\left \langle p_x \right \rangle =p_y=0$, $(\Delta p_x)_{\rm FWHM}=12\ m_ec$ to represent electrons from the first stage. The current of the beam is about 350 amperes. A 2 mm long curved channel with centre profile of $x(mm)=(2-z(mm))^3/(6\times2\times10)$ according to Eq. \ref{eq3} is used to guide laser  \uppercase\expandafter{\romannumeral2}: the curvature varies from $R_0=10$ $\rm mm$ at $s=0$ to infinite at $s=s_1$. The curved channel naturally connects to a 3 mm long straight plasma channel with radius of $30$ $\rm \mu m$. A simulation cell size of 29.4 nm $\times$ 50 nm and 9 particles per cell are chosen. The yellow line in Fig. \ref{fig3}(a) shows the centroid motion of laser  \uppercase\expandafter{\romannumeral1}, which is deflected upward from the boundary of the curved channel. While laser  \uppercase\expandafter{\romannumeral2} is guided to the straight channel with an original tilt incidence angle of $5.7^{\circ}$ and off-axis deviation of $6.33$ $\rm \mu m$. It is found that approximately 12.5\% net enegy of laser  \uppercase\expandafter{\romannumeral2} is lost in the curved channel. In a realistic experiment the length of the curved channel might be extended but the energy of laser  \uppercase\expandafter{\romannumeral2} would have to be increased to compensate for the energy loss. Such an extension also results in weaker transverse laser oscillations. The simulation shows that the main electron beam is well confined by the self-excited wakefields in the plasma and does not lead to large divergence \cite{Thaury2015NC}. However, the head erosion of the beam still diminishes the beam quality and increases the total transverse energy spread \cite{Rosenzweig1991PRA}. As mentioned earlier, in the current scheme the centroid of laser  \uppercase\expandafter{\romannumeral2} oscillates around the tapered equilibrium trajectory $r_{equ}$ and eventually settles down in the straight channel (see the red line in Fig. \ref{fig3}(a)), exciting a stable on-axis wakefield. Electrons passing through the curved channel boundary and injected into the wakefield of laser  \uppercase\expandafter{\romannumeral2} at the entrance of the straight channel. These electrons are then re-accelerated longitudinally with little transverse oscillation due to the reduced transverse kick from the second laser wakefield during the injection process, which also leads to loss of some electrons. Since we only focus on the transition stage in the current study, a simple straight channel with non-tapered density profile is used to re-accelerate. It is found that under such conditions electron dephasing occurs before depletion of laser  \uppercase\expandafter{\romannumeral2} at about z=5 mm with a limited electron energy increase of about 200 MeV. Figures \ref{fig3}(b) and \ref{fig3}(c) show snapshots at injection and dephasing inside the straight channel, respectively. The laser spot and wakefield show no obvious deformation at both times, which suggests that coupling of the two stages is very smooth. Head erosion leads to a loss of about 15\% of electrons during injection and 5\% in the following transverse beam oscillation, which results in finally 80\% electrons remaining in the bubble until dephasing (see the transverse charge distribution indicated by red solid lines in Figs. \ref{fig3}(b,c)). It should be pointed out that because of limitations in available computational resources we have only used a short curved channel. A longer curved plasma channel for realistic experiments would give better laser guiding, higher electron injection rate and smaller transverse oscillation. The second stage could also have a longitudinally tapered plasma density to increase the dephasing length, which would lead to a further increase in the maximum energy gain in the second stage.

\begin{figure}
\includegraphics[width=0.48\textwidth]{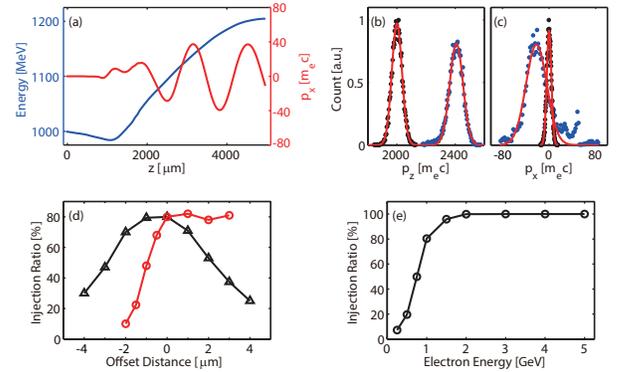}
\caption{
(a) Evolutions of the injected electron beam energy (blue) and transverse momentum (red). (b) and (c) are respectively the initial (black points) and final (blue points) distributions of the electron beam longitudinal and transverse momentum and their Gaussian fitting curves (red). Injection ratio of finally accelerated electrons with different transverse (black) or longitudinal (red) offsets of initial electron beam position (d) and different pre-accelerated electron beam energy(e).
}
\label{fig4}
\end{figure}

We have also studied the electron quality variation and coupling tolerances. The evolution of beam energy (blue line) and transverse momentum (red line) are plotted in Fig. \ref{fig4}(a). Before injection, electrons experience a period of self-propagation and deplete their energy into the background plasma by wakefield excitation. Following trapping in the second accelerator stage, they continuously accelerate until dephasing. However, the transverse momentum of electrons will be resonantly enhanced by the transverse field of the oscillating wake due to the laser centroid oscillation \cite{Cipiccia2011NP}, which is detrimental to high energy acceleration. As discussed above, the curved plasma channel coupler tends to damp the transverse oscillation of laser \uppercase\expandafter{\romannumeral2} after some propagation distance, so that the electron beam transverse momentum $p_x$, does not resonantly increase and remains lower than 40 $m_e c$. Consequently, electrons remain in the bubble until dephasing.

The longitudinal and transverse momenta distributions of the electron beam are plotted in Figs. \ref{fig4}(b) and \ref{fig4}(c). The mean value of $p_z$ increases by more than 400 $m_e c$, while the absolute full width at half maxima (FWHM) of the Gaussian fitting curve increases slightly from 100 $m_e c$ to 129.4 $m_e c$, which corresponds to an increase in the relative energy spread from 5\% to 5.4\%. In contrast, $(\Delta p_x)_{\rm FWHM}$ increases from 12 $m_e c$ to 54.0 $m_e c$. As discussed above, this increase is mainly because of the initial transverse kick of the electrons resulting from the matching of the transverse oscillation of the laser and electron beam, and it can be reduced by using a longer curved plasma channel with a more gradual transition. 

Capture efficiency of finally accelerated electrons with different transverse or longitudinal offsets of the initial electron beam position is illustrated in Fig. \ref{fig4}(d). Transversally, the injection tolerance has a range of about $5$ $\rm \mu m$. Within this range more than half electrons can be trapped in the second stage and be continuously accelerated. Longitudinally, an injection delay variation within $3$ $\rm \mu m$ has no influence on the injection ratio. However, if electrons are injected earlier, then the acceleration gradient and final electron energy decrease due to the improper injection phase in the wakefield. In contrast, if the electron injection is delayed, the injection ratio decreases rapidly. For a further $2$ $\rm \mu m$ delay, almost no electron injection is observed. The injection capture efficiency increases with the pre-accelerated electron energy, as shown in Fig. \ref{fig4}(e). Simulations show that when the initial electron beam energy is higher than 2 GeV, the injection ratio can approach 100\%, which suggests that the coupling scheme would be more efficient in the later stages, and no further modifications of the coupling would be needed. This is particularly advantageous for future multi-staged LWFA based TeV colliders. 

In conclusion, we have shown that using a specially designed plasma channel as a transition stage, a multistage LWFA can be constructed. A curved plasma channel can be used to guide an intense laser pulse into a straight channel, while minimising transverse oscillations and laser profile distortions. The damping of laser pulse oscillation guarantees effective confinement of the injected electron beam in the second stage where it is further accelerated. The pre-accelerated electron beam transverse dispersion is also overcome by self-generated wakefield focusing. A test PIC simulation shows that, with moderate laser and channel parameters, 80\% of electrons with initial 1 GeV energy can be injected into the second stage and an energy gain of 200 MeV is achieved, while almost preserving the electron beam energy spread. We have shown that there are high tolerances on beam transverse and longitudinal positions for injection, which suggests realistic experiments for demonstrating this type of inter-stage coupling. Such kind of curved plasma channel can be made from micromachining using a femtosecond laser \cite{Jaroszynski2006PTRSA}. The excellent properties of this staging method, in particular compactness and weak dependence on the electron beam parameters, make it suitable for future multistage accelerators.

\section*{Acknowledgement}
This work was supported by the National Basic Research Program of China (Grants No. 2013CBA01504), the National Science Foundation of China (Grant No. 11774227, 11374209, 11374210). ZMS and DAJ acknowledge the support from UK Engineering and Physical Sciences Research Council (EPSRC) (Grants No. EP/N028694/1), EC's H2020 EuPRAXIA (Grant No. 653782), LASERLAB-EUROPE (Grant No. 654148) and the Leverhulme Trust. Work at LBNL was supported by the Director, Office of Science, Office of High Energy Physics, of the U.S. Department of Energy under Contract No. DE-AC02-05CH11231.The authors would like to acknowledge the OSIRIS Consortium, consisting of UCLA and IST for the use of OSIRIS and the visXD framework. Simulations were performed on the $\Pi$ Supercomputer at SJTU.

\bibliographystyle{unsrt}

\begin{thebibliography}{40}

\bibitem{Tajima1979PRL}
T.~Tajima, and J.~M.~Dawson,
Phys. Rev. Lett. \textbf{43}, 267 (1979).

\bibitem{Pukhov2002APB}
A.~Pukhov, and J.~Meyer-ter-Vehn,
Appl. Phys. B \textbf{74}, 355(2002).

\bibitem{Esarey2009RMP}
E.~Esarey, C.~B.~Schroeder, and W.~P.~Leemans,
Rev. Mod. Phys. \textbf{81}, 1229 (2009).

\bibitem{Martins2010NP}
S.~F.~Martins, R.~A.~Fonseca, W.~Lu, W.~B.~Mori, and L.~O.~Silva,
Nat. Phys. \textbf{6}, 311 (2010).

\bibitem{He2013APL}
Z.-H.~He, A.~G.~R.~Thomas, B.~Beaurepaire, J.~A.~Nees, B.~Hou, V.~Malka, K.~Krushelnick, and J.~Faure,
Appl. Phys. Lett. \textbf{120}, 064104 (2013).

\bibitem{Karsch2007NJP}
S.~Karsch, J.~Osterhoff, A.~Popp, T.~P.~Rowlands-Rees, Zs~Major, M.~Fuchs, B.~Marx, R.~H\"orlein, K.~Schmid, L.~Veisz, S.~Becker, U.~Schramm, B.~Hidding, G.~Pretzler, D.~Habs, F.~Gr\"uner, F.~Krausz, S.~M.~Hooker,
New J. Phys. \textbf{9}, 415 (2007).

\bibitem{Kim2013PRL}
H.~T.~Kim, K.~H.~Pae, H.~J.~Cha, I.~J.~Kim, T.~J.~Yu, J.~H.~Sung, S~.K.~Lee, T.~M.~Jeong, and J.~Lee,
Phys. Rev. Lett. \textbf{111}, 165002 (2013).

\bibitem{Wang2013NC}
X.~Wang, R.~Zgadzaj, N.~Fazel, Z.~Li, S.~A.~Yi, X.~Zhang, W.~Henderson, Y.-Y.~Chang, R.~Korzekwa, H.-E.~Tsai, C.-H.~Pai, H.~Quevedo, G.~Dyer, E.~Gaul, M.~Martinez, A.~C.~Bernstein, T.~Borger, M.~Spinks, M.~Donovan, V.~Khudik, G.~Shvets, T.~Ditmire, and M.~C.~Downer,
Nat. Commun. \textbf{4}, 1988 (2013).

\bibitem{Corde2013RMP}
S.~Corde, K.~Ta~Phuoc, G.~Lambert, R.~Fitour, V.~Malka, and A.~Rousse,
Rev. Mod. Phys. \textbf{85}, 1 (2013).

\bibitem{Yan2017NP}
W.~C.~Yan, C.~Fruhling, G.~Golovin, D.~Haden, J.~Luo, P.~Zhang, B.~Zhao, J.~Zhang, C.~Liu, M.~Chen, S.~Chen, S.~Banerjee, and D.~Umstadter,
Nat. Photon. \textbf{11}, 514 (2017).

\bibitem{Cipiccia2011NP}
S.~Cipiccia, M.~R.~Islam, B.~Ersfeld, R.~P.~Shanks, E.~Brunetti, G.~Vieux, X.~Yang, R.~C.~Issac, S.~M.~Wiggins, G.~H.~Welsh, M.-P.~Anania, D.~Maneuski, R.~Montgomery, G.~Smith, M.~Hoek, D.~J.~Hamiltion, N.~R.~C.~Lemos, D.~Symes, P.~P.~Rajeev, V.~O.~Shea, J.~M.~Dias, and D.~A.~Jaroszynski,
Nat. Phys. \textbf{7}, 867 (2011).

\bibitem{Luo2016SR}
J.~Luo, M.~Chen, M.~Zeng, J.~Vieira, L.~L.~Yu, S.~M.~Weng, L.~O.~Silva, D.~A.~Jaroszynski, Z.~M.~Sheng, and J.~Zhang,
Sci. Rep. \textbf{6}, 29101 (2016).

\bibitem{Leemans2009PT}
W.~Leemans, and E.~Esarey,
Phys. Today \textbf{62}, 44 (2009).

\bibitem{Schroeder2010PRSTAB}
C.~B.~Schroeder, E.~Esarey, C.~G.~R.~Geddes, C.~Benedetti, and W.~P.~Leemans,
Phys. Rev. Special Topics Accel. Beams \textbf{13}, 101301 (2010).

\bibitem{Leemans2014PRL}
W.~P.~Leemans, A.~J.~Gonsalves, H.-S.~Mao, K.~Nakamura, C.~Benedetti, C.~B.~Schroeder, Cs.~T\'oth, J.~Daniels, D.~E.~Mittelberger, S.~S.~Bulanov, J.-L.~Vay, C.~G.~R.~Geddes, and E.~Esarey,
Phys. Rev. Lett. \textbf{113}, 245002 (2014).

\bibitem{Lu2007PRSTAB}
W.~Lu, M.~Tzoufras, C.~Joshi, F.~S.~Tsung, W.~B.~Mori, J.~Vieira, R.~A.~Fonseca, and L.~O.~Silva,
Phys. Rev. Special Topics Accel. Beams \textbf{10}, 061301 (2007).

\bibitem{Kneip2009PRL}
S.~Kneip, S.~R.~Nagel, S.~F.~Martins, S.~P.~D.~Mangles, C.~Bellei, O.~Chekhlov, R.~J.~Clarke, N.~Delerue, E.~J.~Divall, G.~Doucas, K.~Ertel, F.~Fiuza, R.~Fonseca, P.~Foster, S.~J.~Hawkes, C.~J.~Hooker, K.~Krushelnick, W.~B.~Mori, C.~A.~J.~Palmer, K.~Ta~Phuoc, P.~P.~Rajeev, J.~Schreiber, M.~J.~V.~Streeter, D.~Urner, J.~Vieira, L.~O.~Silva, and Z.~Najmudin,
Phys. Rev. Lett. \textbf{103}, 035002 (2009).

\bibitem{Clayton2010PRL}
C.~E.~Clayton, J.~E.~Ralph, F.~Albert, R.~A.~Fonseca, S.~H.~Glenzer, C.~Joshi, W.~Lu, K.~A.~Marsh, S.~F.~Martins, W.~B.~Mori, A.~Pak, F.~S.~Tsung, B.~B.~Pollock, J.~S.~Ross, L.~O.~Silva, and D.~H.~Froula,
Phys. Rev. Lett. \textbf{105}, 105003 (2010).

\bibitem{Liu2011PRL}
J.~S.~Liu, C.~Q.~Xia, W.~T.~Wang, H.~Y.~Lu, Ch.~Wang, A.~H.~Deng, W.~T.~Li, H.~Zhang, X.~Y.~Liang, Y.~X.~Leng, X.~M.~Lu, C.~Wang, J.~Z.~Wang, K.~Nakajima, R.~X.~Li, and Z.~Z.~Xu,
Phys. Rev. Lett. \textbf{107}, 035001 (2011).

\bibitem{Guillaume2015PRL}
E.~Guillaume, A.~D\"opp, C.~Thaury, K.~Ta~Phuoc, A.~Lifschitz, G.~Grittani, J.-P.~Goddet, A.~Tafzi, S.~W.~Chou, L.~Veisz, and V.~Malka,
Phys. Rev. Lett. \textbf{115}, 155002 (2015).

\bibitem{Mirzaie2015SR}
M.~Mirzaie, S.~Li, M.~Zeng, N.~A.~M.~Hafz, M.~Chen, G.~Y.~Li, Q.~J.~Zhu, H.~Liao, T.~Sokollik, F.~Liu, Y.~Y.~Ma, L.~M.~Chen, Z.~M.~Sheng, and J.~Zhang,
Sci. Rep. \textbf{5}, 14659 (2015).

\bibitem{Steinke2016N}
S.~Steinke, J.~van~Tilborg, C.~Benedetti, C.~G.~R.~Geddes, C.~B.~Schroeder, J.~Daniels, K.~K.~Swanson, A.~J.~Gonsalves, K.~Nakamura, N.~H.~Matlis, B.~H.~Shaw, E.~Esarey, and W.~P.~Leemans,
Nature (London) \textbf{530}, 190 (2016).

\bibitem{Sokollik2010AIPCP}
T.~Sokollik, S.~Shiraishi, J.~Osterhoff, E.~Evans, A.~J.~Gonsalves, K.~Nakamura, J.~van~Tilborg, C.~Lin, Cs.~T\'oth, and W.~P.~Leemans,
AIP Conf. Proc. \textbf{1299}, 233 (2010).

\bibitem{Tilborg2015PRL}
J.~van~Tilborg, S.~Steinke, C.~G.~R.~Geddes, N.~H.~Matlis, B.~H.~Shaw, A.~J.~Gonsalves, J.~V.~Huijts, K.~Nakamura, J.~Daniels, C.~B.~Schroeder, C.~Benedetti, E.~Esarey, S.~S.~Bulanov, N.~A.~Bobrova, P.~V.~Sasorov, and W.~P.~Leemans,
Phys. Rev. Lett. \textbf{115}, 184802 (2015).

\bibitem{Ehrlich1996PRL}
Y.~Ehrlich, C.~Cohen, A.~Zigler, J.~Krall, P.~Sprangle, and E.~Esarey,
Phys. Rev. Lett. \textbf{77}, 4186 (1996).

\bibitem{Reitsma2007POP}
A.~J.~W.~Reitsma and D.~A.~Jaroszynski, 
Phys. Plasmas \textbf{14}, 053104 (2007).

\bibitem{Chen2016LSA}
M.~Chen, J.~Luo, F.~Y.~Li, F.~Liu, Z.~M.~Sheng, and J.~Zhang,
Light Sci. Appl. \textbf{5}, e16015 (2016).

\bibitem{Palastro2017POP}
J.~P.~Palastro, D.~Kaganovich, B.~Hafizi, Y.-H.~Chen, L.~A.~Johnson, J.~R.~Pe$\rm \tilde{n}$ano, M.~H.~Helle, and A.~A.~Mamonau,
Phys. Plasmas \textbf{24}, 033119 (2017).

\bibitem{Hooker2007P}
S.~M.~Hooker, A.~J.~Gonsalves, D.~A.~Jaroszynski, W.~P.~Leemans, WO/2008/032050 ''Charged particle accelerator and radiation source'', Date of Patent: 11.09.2007.

\bibitem{Fonseca2002LNCSE}
R.~A.~Fonseca, L.~O.~Silva, F.~S.~Tsung, V.~K.~Decyk, W.~Lu, C.~Ren, W.~B.~Mori, S.~Deng, S.~Lee, T.~Katsouleas, and J.~C.~Adam,
Lect. Notes Comput. Sci. \textbf{2331}, 342 (2002).

\bibitem{Thaury2015NC}
C.~Thaury, E.~Guillaume, A.~D\"opp, R.~Lehe, A.~Lifschitz, K.~Ta~Phuoc, J.~Gautier, J-P~Goddet, A.~Tafzi, A.~Flacco, F.~Tissandier, S.~Sebban, A.~Rousse, and V.~Malka,
Nat. Commun. \textbf{6}, 6860 (2015).

\bibitem{Rosenzweig1991PRA}
J.~B.~Rosenzweig, B.~Breizman, T.~Katsouleas, and J.~J.~Su,
Phys. Rev. A \textbf{44}, R6189 (1991).

\bibitem{Jaroszynski2006PTRSA}
D.~A.~Jaroszynski, R.~Bingham, E.~Brunetti, B.~Ersfeld, J.~Gallacher, B.~van~der~Geer, R.~Issac, S.~P.~Jamison, D.~Jones, M.~de~Loos, A.~Lyachev, V.~Pavolv, A.~Reitsma, Y.~Saveliev, G.~Vieux, and S.~M.~Wiggins,
Phil. Trans. R. Soc. A. \textbf{364}, 689 (2006).

\end{thebibliography}

\end{document}